\documentclass{article}

\usepackage{microtype}
\usepackage{graphicx}
\usepackage{subfigure}
\usepackage{booktabs} 

\usepackage{hyperref}
\usepackage[capitalise,noabbrev,nameinlink]{cleveref}

\newcommand{\pf}{prefill}
\newcommand{\toolname}{AcceLLM}
\newcommand{\sota}{state-of-the-art}

\usepackage[accepted]{mlsys2025}

\mlsystitlerunning{AcceLLM: Accelerating LLM Inference using Redundancy for Load Balancing and Data Locality}

\begin{document}

\twocolumn[
\mlsystitle{
AcceLLM: Accelerating LLM Inference using Redundancy for Load Balancing and Data Locality
}

\vspace{-0.5cm}


\mlsyssetsymbol{equal}{*}

\begin{mlsysauthorlist}
\mlsysauthor{Ilias Bournias}{hu,equal}
\mlsysauthor{Lukas Cavigelli}{hu}
\mlsysauthor{Georgios Zacharopoulos}{hu,equal}
\end{mlsysauthorlist}

\mlsysaffiliation{hu}{Computing Systems Lab, Huawei, Zurich Research Center}


\mlsyskeywords{Machine Learning, MLSys, LLMs, Load Balancing, Data Locality}

\vskip 0.3in

\vspace{-0.5cm}

\thispagestyle{plain}
\pagestyle{plain}

\begin{abstract}
Large Language Model (LLM) inference on large-scale systems is expected to dominate future cloud infrastructures. 
Efficient LLM inference in cloud environments with numerous AI accelerators is challenging, necessitating extensive optimizations for optimal performance. Current systems batch prefill and decoding to boost throughput but encounter latency issues, while others disaggregate these phases, leading to resource underutilization.
We propose AcceLLM, a novel method addressing latency and load balancing, inspired by the cache data management. It strategically utilizes  redundant data to enhance inference via load balancing and optimal hardware use. Simulated evaluations on Nvidia H100 GPU and Huawei Ascend 910B2 show AcceLLM surpasses state-of-the-art systems up to 30\% in latency and efficiency, handling diverse workloads effectively.
\end{abstract}
]


\printAffiliationsAndNotice{\mlsysEqualContribution}

\section{Introduction}
\label{sec:intro}

The surge in demand for Large Language Model (LLM) inference is driven by the widespread adoption of Natural Language Processing (NLP) applications across industries, including chatbots, virtual assistants, analytics, and automated content creation. This growth \cite{schillaci2024llm} is expected to further strain the hardware resources of cloud infrastructures.

Leading tech companies such as OpenAI, Microsoft, and Google utilize extensive GPU and TPU resources to handle LLM inference tasks \cite{llm_inf_cost}, which involve two primary computational stages: prefill and decoding. Prefill processes the initial prompt, while decoding generates tokens sequentially, employing multi-head self-attention and feed-forward networks.

In a heterogeneous computing ecosystem, exemplified by a cloud infrastructure, 
high-speed interconnects enable efficient collaboration among diverse processors \cite{mccoll2017superclouds}.
Software mechanisms facilitate workload distribution and coordination across different levels, from hardware acceleration to task scheduling and load balancing.

However, current LLM inference systems experience latency spikes between output token generation \cite{fastertransformer,2023Pageattention,2023sarathi,2024blockllm,2024hetegen,2023spotserve,2023flexgen,2024orion}, adversely affecting user experience and necessitating over-provisioning of resources. To address this, researchers have explored separating prefill and decoding stages across different machines, or \emph{instances} that can employ multiple devices \cite{2024Tetrinfer, 2023splitwise, 2024distserve} and involve transferring a large state, often referred to as KV cache.

\begin{figure}[t]
\centering
\includegraphics[width=\linewidth]{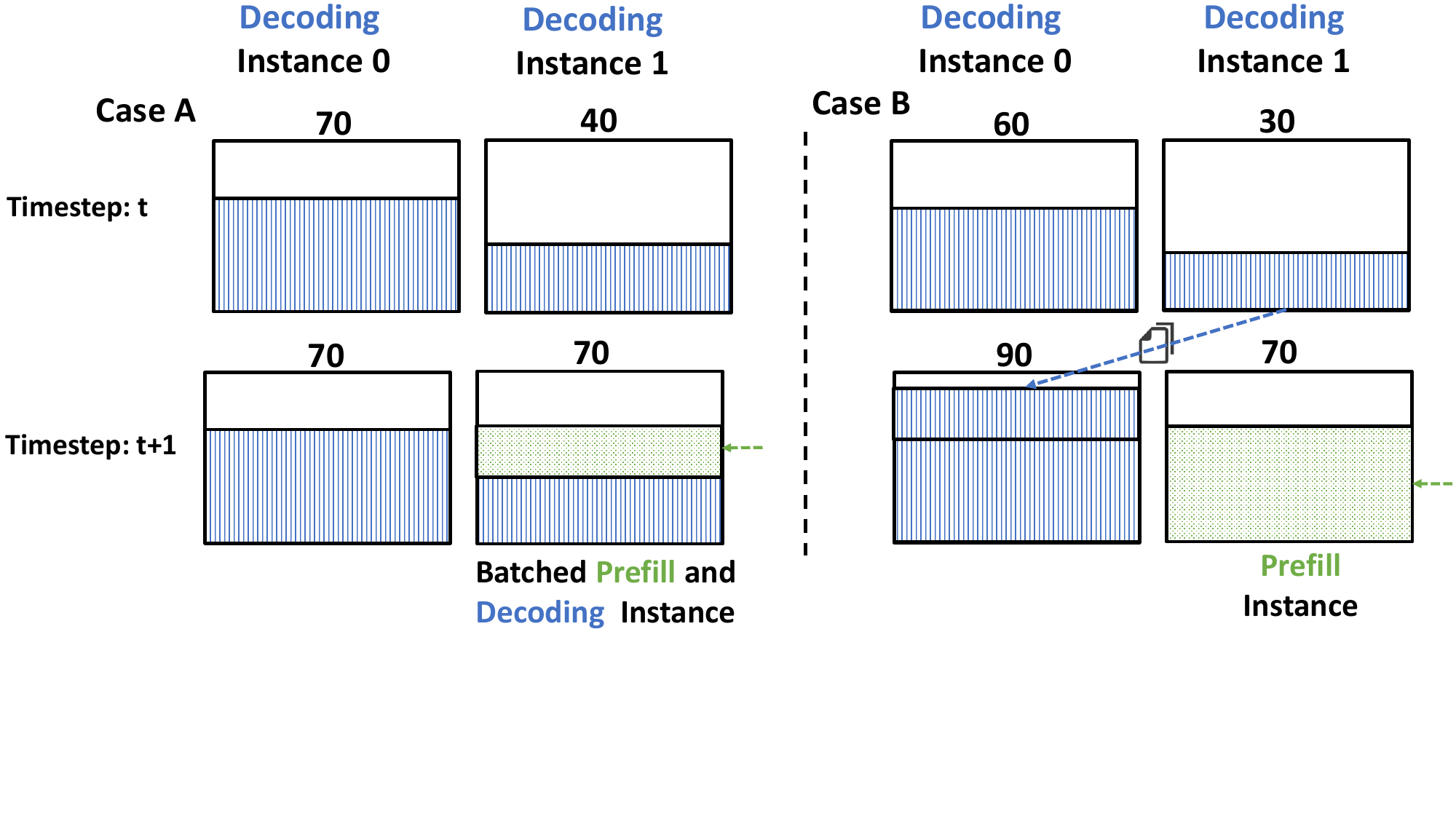}
\vspace{-1.7cm}
\caption{Peak Latency Challenge. Case A: Batched Prefill and Decoding resulting in a latency peak.
Case B: Decoding data transfer overhead between instances due to a large prefill request.
}
\vspace{-0.7cm}
\label{fig:peak_latency}
\end{figure}

This approach aims to manage diverse task granularities and scale resource pools according to computational demands, mitigating latency peaks caused by lengthy prefill tasks interleaved with token generation. 
It also puts away with latency peaks that otherwise occur when the commonly much longer prefill tasks are executed between single token decoding/generation tasks. We visualize this challenge in \cref{fig:peak_latency} (Case A), where an incoming prefill request necessitates batching and simultaneous computation alongside a decoding task on the same computing instance, resulting in increased latency. 

\begin{figure}[t]
\centering
\includegraphics[width=1.05\linewidth]
{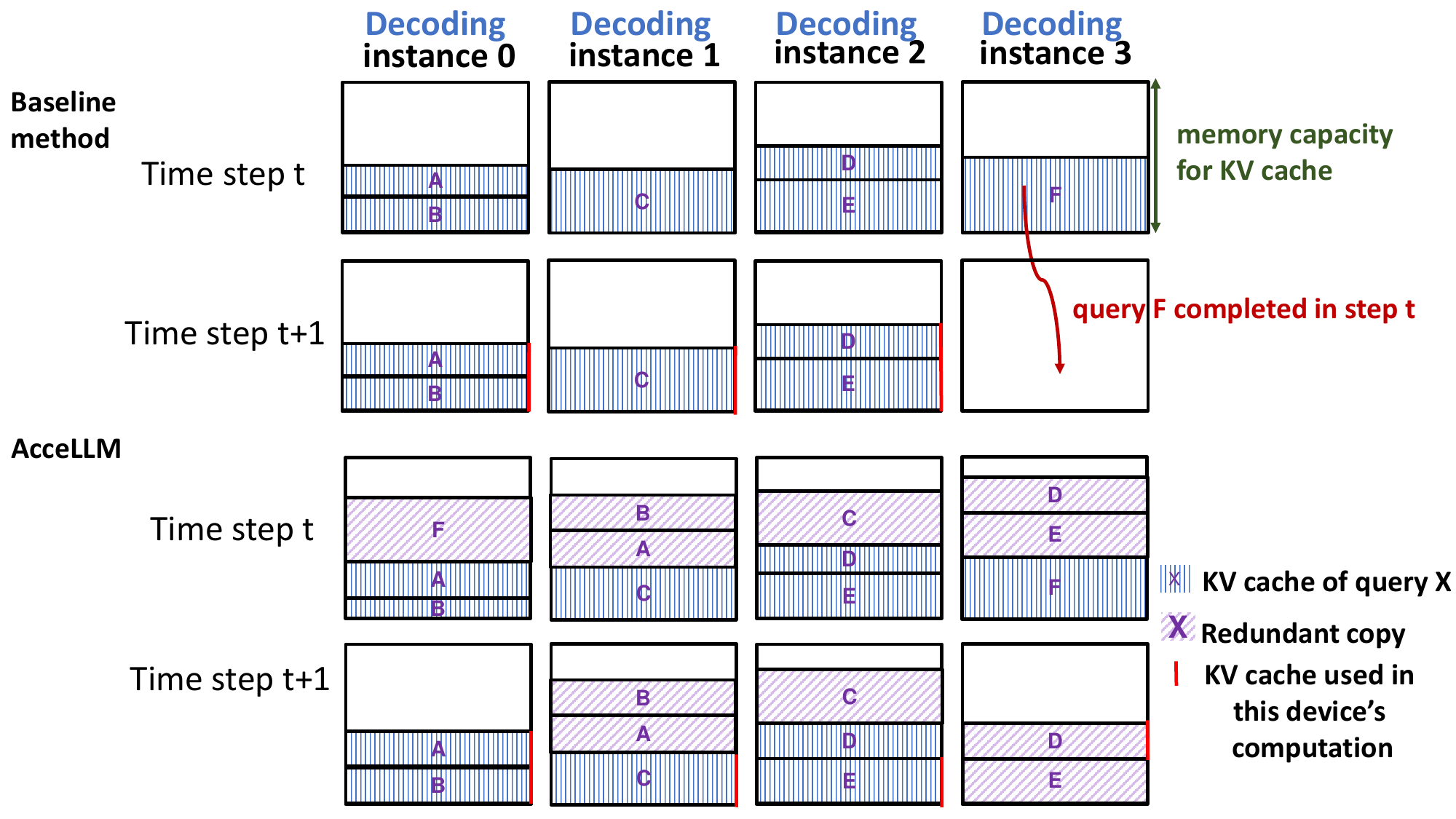}
\vspace{-0.8cm}
\caption{
The KV cache of different queries is allocated on different instances.
Query F (KV cache on Instance 3), is completed in time step $t$. In step $t+1$ (and many later), device 3 is idle.
We use the free space on the devices to keep redundant copies of the KV cache. 
After query F is completed, there is enough KV cache on every device to keep good \textbf{load balancing}.
}
\vspace{-0.4cm}
\label{fig:load_balancing}
\end{figure}

However, the methods mentioned above \emph{do not} tackle two main challenges that can significantly hinder performance: a) prefill-related latency peaks and b) lack of load balancing. The first is due to the
inherent need to copy the LLM query's state (KV cache) before the decoding nodes can start producing output tokens, which in turn adds respective communication cost, as seen in \cref{fig:peak_latency} (Case B). 
Furthermore, they fail to tackle challenges associated with load imbalances concerning both prefill and decoding requests across devices, leading to underutilization of available hardware resources, as seen in \cref{fig:load_balancing}.

To overcome these challenges and enhance hardware resource utilization, we propose \toolname, a novel method offering data locality, load balancing, and dynamic instances by employing redundant data copies across computing instances.
We provide the following contributions: 
\vspace{-0.4cm}
\begin{enumerate}
    
    \item 
    We propose  a novel method ---\toolname--- leveraging \textbf{redundant data} in the form of KV cache copies across instances of multiple devices, drawing inspiration from the data management in caches and cloned computing. 
    
    \item We show how AcceLLM reduces latency and provides superior cost efficiency by achieving effective \textbf{load balancing} among nodes.  
    
    \item We show an \textbf{improved utilization} as instances can serve both, generation or prefill tasks, dynamically according to the workload demand, without requiring any internal adjustments of the LLM's weights.

    \item We offer an \textbf{extensive evaluation} of AcceLLM on two leading simulated platforms, Nvidia's H100 GPU and Huawei's Ascend 910B2, and compare it against \sota\ methods.

\end{enumerate}

\vspace{-0.5cm }
\section{Related Work}
\label{sec:rw}

The exceptionally high demand for serving LLM inference requests has lead to an increasing number of LLM serving systems.
Conventional request-level batching systems, such as Faster-Transformer \cite{fastertransformer}, employ a decode-prioritizing scheduling approach. In these systems, a batch of requests is processed by first completing the prefill phase for all requests, followed by scheduling their decode phase. The batch is considered complete only after all requests within it have finished decoding. While this method optimizes the time between tokens, it significantly increases the queuing time for incoming requests, as they must wait for the current batch to complete before execution. Additionally, it reduces throughput since batch execution continues until the last request is finished, even if some requests complete earlier. 

Orca~\cite{2022Orca} tackles this challenge by employing iteration-level scheduling. They optimize GPU resource utilization by batching the prefill and decoding phases of different requests in different phases. This approach minimizes redundant usage of High Bandwidth Memory (HBM). Similarly, Sarathi \cite{2023sarathi} and DeepSpeed \cite{2024deepspeed} adopt analogous strategies, but with slight variations. 
Sarathi chunks the prefill phase to align with the decoding phase, aiming to maximize GPU resource utilization and make it compute bound. This approach ensures uniform compute iterations, effectively minimizing pipeline bubbles. DeepSpeed further refines this by decomposing long prompts into smaller segments and scheduling them across multiple forward passes (iterations), with only the final pass performing generation. 

Sarathi-Serve \cite{2024Sarathi-Serve} utilizes chunked-prefills from Sarathi \cite{2023sarathi} for uninterrupted scheduling, enabling new batch additions without stalling ongoing decodes. This stall-free scheduling supports larger batch sizes and improves throughput while controlling latency. Vanilla LLM (vLLM) \cite{2023Pageattention} optimizes GPU memory by segmenting the KV cache by sequence length but prioritizes prefill, which increases decode latency. Infinite-LLM \cite{2024InfiniteLLM} employs DistAttention, a distributed attention algorithm, and proposes DistKV-LLM to manage KV caches across GPU and CPU memories in the data center. However, token generation is slowed as single requests may be partially served by multiple instances.

LoongServer \cite{2024loongserve} introduces elastic sequence parallelism by employing elastic scaling mechanisms and a scalable scheduling algorithm, dynamically adjusting resource allocation based on workload demands. It focuses on efficiently handling very long input sequences by flexibly managing the KV cache across various phases. However, like Infinite-LLM, it faces challenges related to data locality. Despite their distinct optimizations, these approaches batch the prefill and decode phases together, leading to reduced throughput and latency spikes during token generation.

 To address the challenges of batching the prefill and decode phases, recent methods propose using separate instances for each phase. Splitwise \cite{2023splitwise} optimizes token generation by transferring KV caches from prefill to decode instances via high-speed GPU interconnects, employing a two-level scheduling system: a cluster-level scheduler assigns requests to machines, and a machine-level scheduler manages batching within each machine. Similarly, TetriInfer \cite{2024Tetrinfer} segments prompts into fixed-size chunks to maintain high compute efficiency and uses a two-level scheduling algorithm to reduce decode-phase bottlenecks. Additionally, it employs an LLM-based classifier, or "predict model," to categorize token lengths, enhancing resource allocation efficiency for each request.

 DistServe \cite{2024distserve} allocates prefill and decoding tasks to separate GPUs, optimizing resource allocation and parallelism based on time-to-first-token and time-per-token requirements while exploring varied batching strategies. It periodically reallocates resources according to workload patterns, monitored by a profiler that tracks input/output lengths and arrival times. 
 
 DejaVu \cite{2024Dejavu} enables efficient KV cache streaming, disaggregating prefill and decode phases through micro-batching and KV cache swapping between GPU and CPU, reducing GPU memory demands and supporting larger batch sizes on limited hardware. To ensure fault tolerance, it duplicates KV caches in persistent or remote CPU memory, allowing restoration and resumption of inference after failures. Most methods dedicate instances to either prefill or decode tasks, without load balancing between them, which limits request throughput.

 \section{Generative LLM Inference} \label{sec:GenInf}

Generative Large Language Model (LLM) inference has two main stages: prefill and decoding. In prefill, the model processes the user's prompt to generate the first token. Then, in decoding, the model iteratively generates tokens, each feeding into the next, until it hits the maximum token limit, a user-defined stop criterion, or produces an End Of Sentence (EOS) token, signaling the end of generation.

\subsection{The Transformer Architecture}

The Transformer block is the fundamental unit in LLM architectures. It includes the QKV Linear layer, which projects input data into queries (Q), keys (K), and values (V)\cite{2023surveytransformer}. These projections feed into the multi-head self-attention mechanism, the main module in the Transformer block, allowing the model to focus on different parts of the input sequence with varying importance. Multi-head attention enables the model to process information from multiple subspaces concurrently. Following this is the Feed-Forward Neural Network (FFN), typically a three-layer network with an activation function in between. This general structure is consistent across Transformer implementations.

\subsection{Prefill}

\begin{figure}
  \centering
  \includegraphics[width=0.47\linewidth]{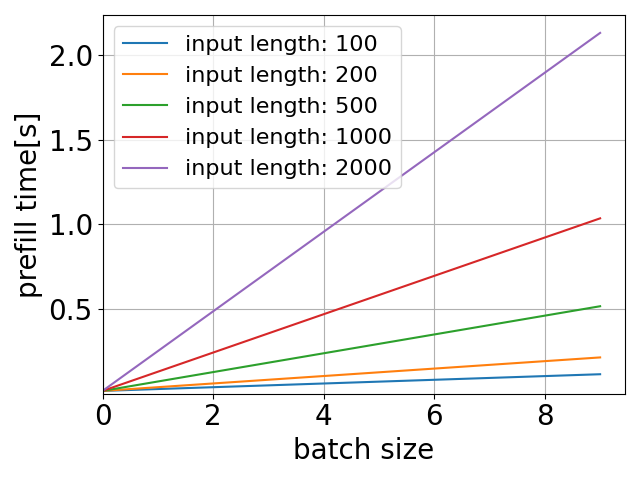}
  \hfill
  \includegraphics[width=0.47\linewidth]{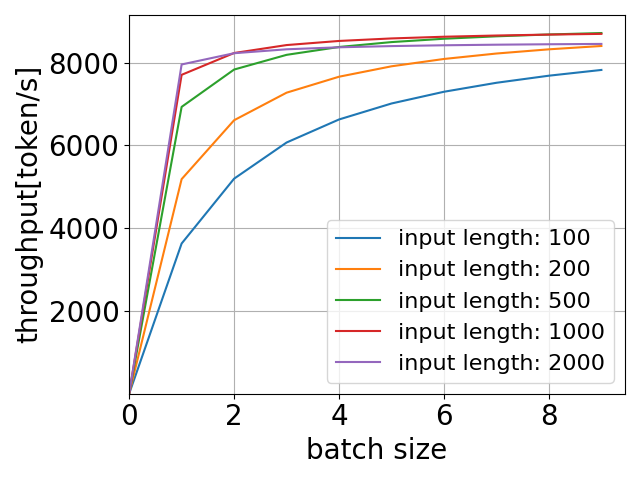}
  \vspace{-0.4cm}
  \caption{Prefill-phase execution time and throughput.}
\vspace{-0.5cm}
  \label{fig:prefill}
\end{figure}

Prefill constitutes a computationally demanding task \cite{2023splitwise}, constrained by the processing capacity of GPU devices. \cref{fig:prefill} illustrates both, the completion time of the prefill process showcasing its dependency on both prompt length and batching, and the number of generated tokens, also contingent on prompt length and batching. The throughput increases while batching more requests, until a certain plateau is reached as the compute operations outweigh the data transfers. To further increase throughput, parallelization must be employed, assigning batches of requests to different instances.

\begin{figure}
  \centering
  \includegraphics[width=0.47\linewidth]{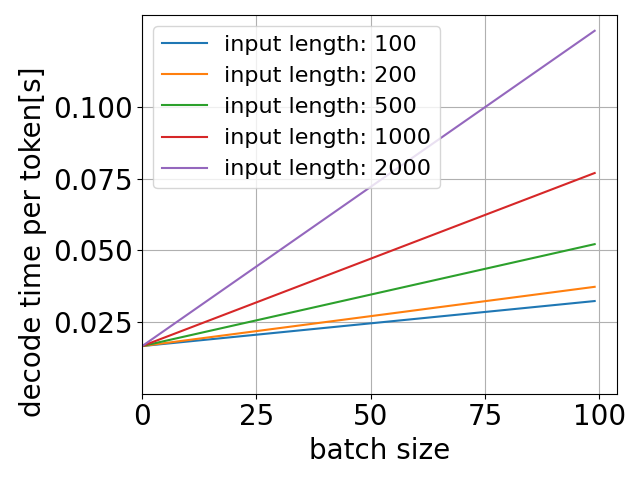}
  \hfill
  \includegraphics[width=0.47\linewidth]{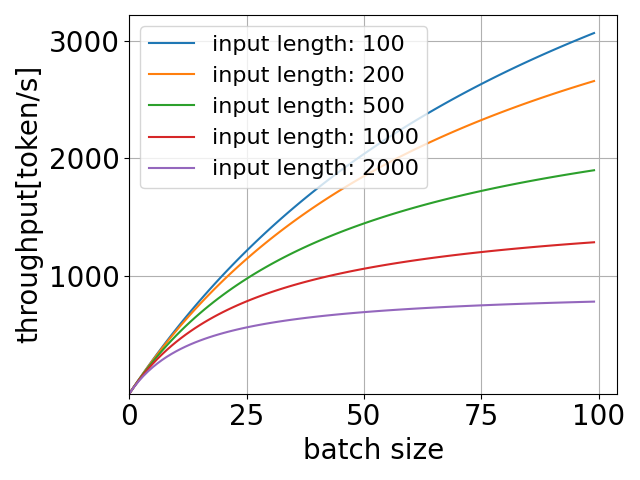}
   \vspace{-0.5cm}
  \caption{Decoding-phase execution time and throughput.}
   \vspace{-0.4cm}
  \label{fig:decoding}
\end{figure}

\subsection{Decoding}

Contrary to prefill, decoding is bounded by the HBM memory bandwidth as the amount of computation is minimal compared to the data transfer times from the memory to the compute units \cite{2024hongflashdecoding}. In order to make better allocation of the GPU resources it is essential to apply batching between the decoding requests since it allows re-use of the LLM's weights for multiple requests. In \cref{fig:decoding}, we show the computation time and the token generation throughput for batched decoding. 

As batch sizes increase, throughput rises due to fewer memory reads \cite{2022Orca}. Initially, this is independent of input length and with larger batch sizes, throughput stabilizes at distinct values for each input length. For longer inputs, the primary bottleneck becomes waiting for the loading of KV cache, as weight loading time that is independent of the input length becomes irrelevant due to the batching. In this important scenario, the compute time is largely proportional to the number of tokens. 

Notably, this throughput limitation by the HBM bandwidth loading of the KV cache during decoding is also the reason why moving the KV cache from one device to another leads to a latency peak---simply as inter-device interconnect speeds are usually an order of magnitude lower than the local HBM bandwidth. 

\subsection{Target Metrics}

We optimize four key metrics for serving LLM inference. The first, \emph{Time to First Token (TTFT)}, measures the delay in generating the first token after receiving a prompt, critical for applications needing fast initial responses. The second, \emph{Time Between Tokens (TBT)}, tracks token generation throughput, essential for responsive streaming. \emph{Job Completion Time (JCT)} gauges the total time from prompt receipt to End-of-Sequence (EOS), assessing overall inference performance. Finally, \emph{cost efficiency} is defined as throughput per instance, assuming cost correlates with runtime and instance count. While latency metrics (TTFT, TBT, JCT) reflect user experience, \emph{cost efficiency} indicates user expense.

\subsection{Challenges}

\subsubsection{Interference of Prefill and Decoding} \label{subsec:interfe}

Most recent \sota\ inference systems \cite{2023sarathi,2022Orca,2024deepspeed} are batching prefill and decoding requests to enhance token generation throughput and fully utilize GPU computation capabilities. While this strategy maintains a relatively consistent TTFT compared to solely processing prefill requests, it results in a notable increase in TBT, as depicted in \cref{fig:challenges}.

This elevation in TBT occurs due to latency spikes caused by the arrival of one or more new requests which are in the prefill phase while an instance is engaged in decoding tasks. The spikes disrupt the smooth processing flow, causing delays in token generation for ongoing tasks. Consequently, this can lead to an increased JCT for each request.

\subsubsection{Load Imbalance}

The size of prompts and the quantity of generated tokens during the decoding phase are not predetermined for each request. Moreover, they vary significantly between requests. Consequently, there are considerable differences in the number of requests (batch size) handled by individual instances since some requests may complete earlier while others might take much longer time due to their decode phase.  

\begin{figure}
  \centering
 \includegraphics[width=0.49\linewidth]{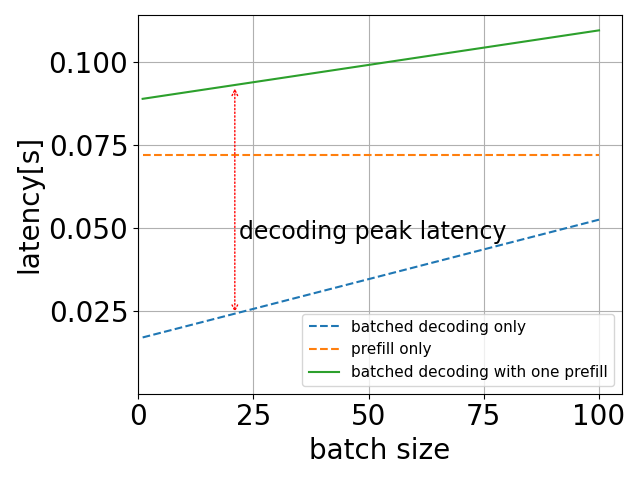}\label{fig:peak_lat}
  \hfill
  \includegraphics[width=0.49\linewidth]{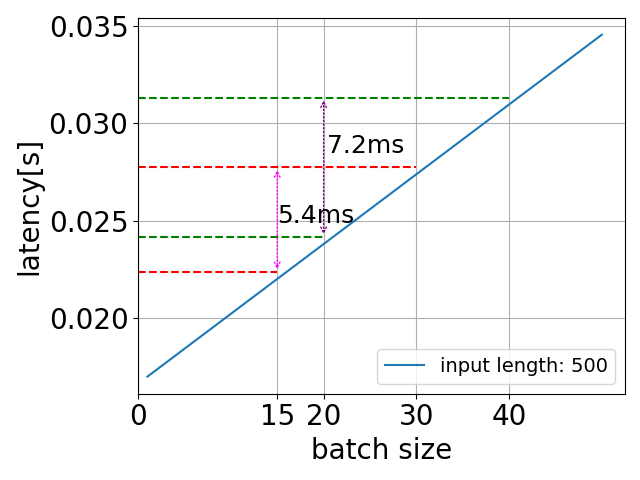}\label{fig:imba}
  \vspace{-0.5cm}
  \caption{
  Left: 
Integrating batching with prefill in the decoding phase increases token generation latency by over  300$\%$. Right: Imbalance arises when batching 40 requests per instance, increasing token generation by 7.2ms compared to parallel execution of the same requests across two instances with a batch size of 20.
  }
  \vspace{-0.3cm}
  \label{fig:challenges}
\end{figure}

The wide variance in request sizes poses a challenge when deploying multiple instances, potentially resulting in an uneven distribution of workload. This discrepancy arises because some instances, constrained by HBM bandwidth, may remain underutilized, leading to wasted hardware resources. Conversely, other instances may become overutilized, failing to leverage available resources effectively. These disparities, largely stemming from variations in batch sizes among instances, contribute to significant fluctuations in TBT (\cref{fig:challenges}) and ultimately result in an increased JCT. Addressing this imbalance is crucial for optimizing system performance and ensuring efficient resource utilization across all instances.

\begin{figure}
  \centering
 \includegraphics[width=\linewidth]{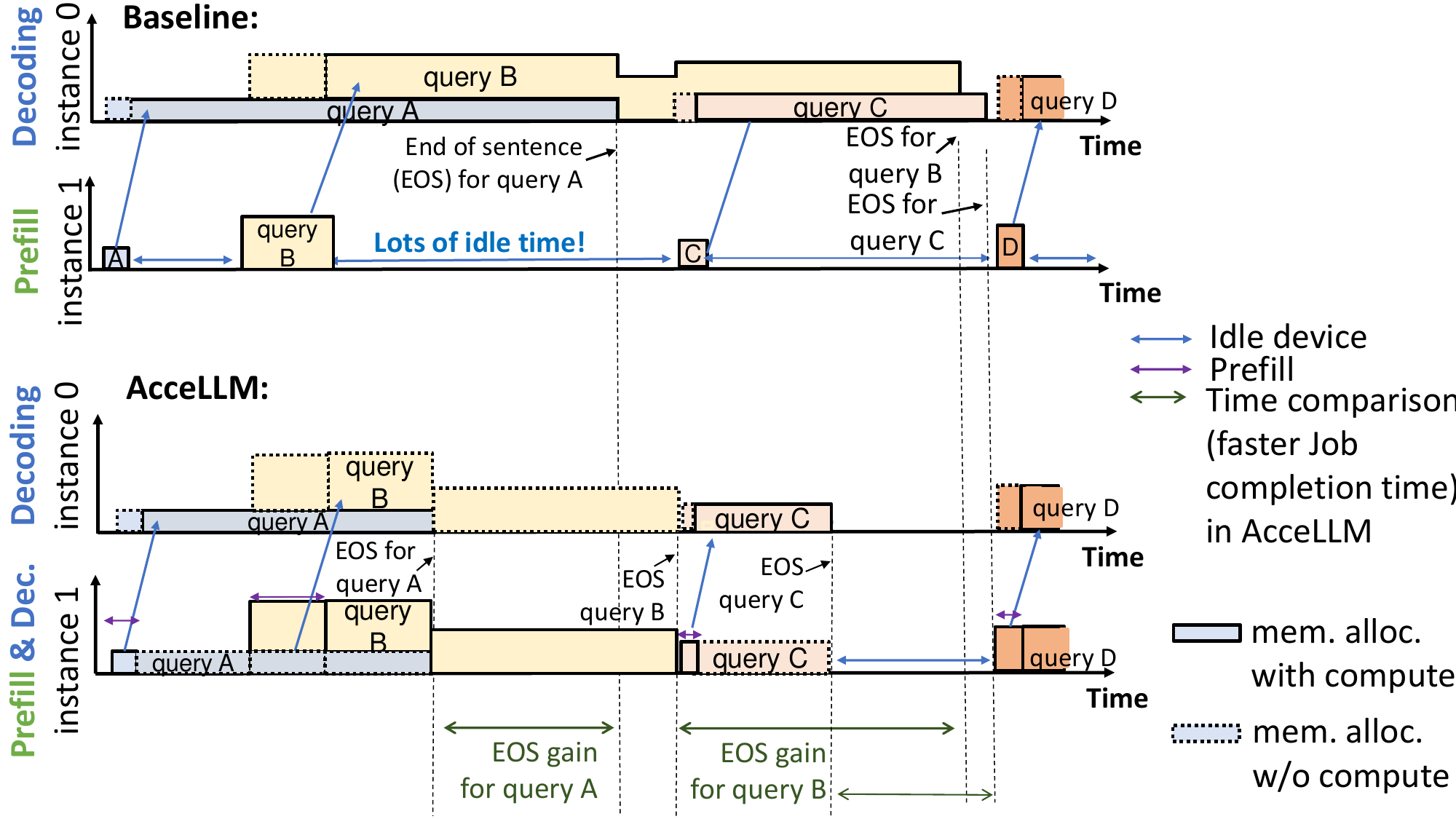}
 \vspace{-0.8cm}
  \caption{The new requests arrive at the same time in both baseline and AcceLLM.
Baseline: the prefill device is idling a lot.
 AcceLLM: no device is idle at any time leading to better load balancing and as a result to strictly better (never worse) performance.
}
  \vspace{-0.5cm}
 \label{fig:idle-load_balance}
\end{figure}

\subsubsection{Dynamic Workload Handling} 

The varying rate at which new requests arrive makes it extremely challenging to allocate the available resources optimally in advance \cite{2024andes,2024melange,2024Tetrinfer}. Certain \sota\  systems \cite{2024andes,2024melange,2024preble}, based on the incoming prompt and a prediction model, decide about the amount of hardware resources that should be allocated. In some other approaches, where disaggregation is applied, the number of instances responsible for computing the prefill and the decode phase is predetermined \cite{2024distserve,2024Tetrinfer,2024Dejavu,2023splitwise} and it solely depends on the incoming request rate. In some of these approaches, \cite{2023splitwise,2024Tetrinfer} there are some extra static instances which can be used and serve either requests that are in the prefill phase or requests which are in the token generation phase depending on the incoming workload. 

However, to the best of our knowledge, \textbf{none of the previous \sota\ works have instances which can be redirected from one phase to the other} \cite{2024Dejavu,2023splitwise,2024distserve,2024Tetrinfer}, depending on the workload and the \emph{currently} served requests in the decode phase. For redirecting one instance, first serving all the requests stored inside, or moving the KV caches stored in it to other instances is required \cite{2024preble}. These approaches result in instances being idle for long periods of time (\cref{fig:idle-load_balance}) when no new requests arrive, or when the number of requests in the decode phase is very low, leading to the increase of TTFT, TBT and JCT. Furthermore, the migration of whole KV caches during decoding for the conversion, also leads to large latency peaks during requests serving making this approach impractical.

\subsection{LLM Inference Models}
Three computation models for LLM inference are investigated within the scope of this paper, that are specifically tailored for large-scale systems.
A) Disaggregation model. Models that disaggregate the \pf\ and the decoding stages \cite{2024Tetrinfer, 2023splitwise, 2024distserve} by employing separate instances dedicated to each stage. 
B) Load balancing model. Models that target equivalent distribution of workloads across all available resources, so that both better utilization and decreased latency can be achieved, as also demonstrated in \cref{fig:imba}.
C) Redundancy model. Models that generate and hold redundant data that can be exploited during both the \pf\ and the decoding stages of LLM inference.
We propose \textbf{\toolname}, a model that aims at achieving data locality and load balancing by using redundancy strategically in order to eliminate unnecessary communication among instances.

\subsubsection{Disaggregation Model}
In order to address the peak latency problem presented in \cref{fig:peak_latency} (\cref{sec:intro}), models that disagregate the \pf\ and decoding stages have been suggested \cite{2024Tetrinfer, 2023splitwise, 2024distserve}. 
This approach successfully tackles the latency resulting from a batched \pf\ and decoding instance, as seen in Case A (\cref{fig:peak_latency}), however it does not address Case B where there can be latency due to data transfer overhead among instances.

\subsubsection{Load Balancing Model}
In order to address load imbalance challenges, as mentioned in the previous section, a balanced workload distribution is suggested by our method. Moreover, as seen in \cref{fig:challenges}, a large batch size handled by a single instance, for instance batch size 40, would result into a larger latency equal to 7.2ms compared to two instances of batch size equal to 20 for any input length. Therefore, a balanced distribution of allocated resources can lead to faster inference times.

\subsubsection{Redundancy Model}
In order to address all the challenges mentioned above, we introduce a novel concept of performing LLM inference by generating redundant data across instances, inspired from the data management in caches. This model and our method is explained in detail in the following section.

\section{
\toolname\
\label{sec:accellm}
}

\subsection{Methodology}
Drawing upon the insights outlined in \cref{sec:GenInf}, we introduce AcceLLM, an LLM inference system leveraging redundant KV cache data to split prefill and decoding requests. This approach aims at minimizing latency through load balancing and the utilization of flexible instances capable of serving either for generation or prefill tasks, depending on the workload.

\begin{figure}[t]
\centering
\includegraphics[width=\linewidth]{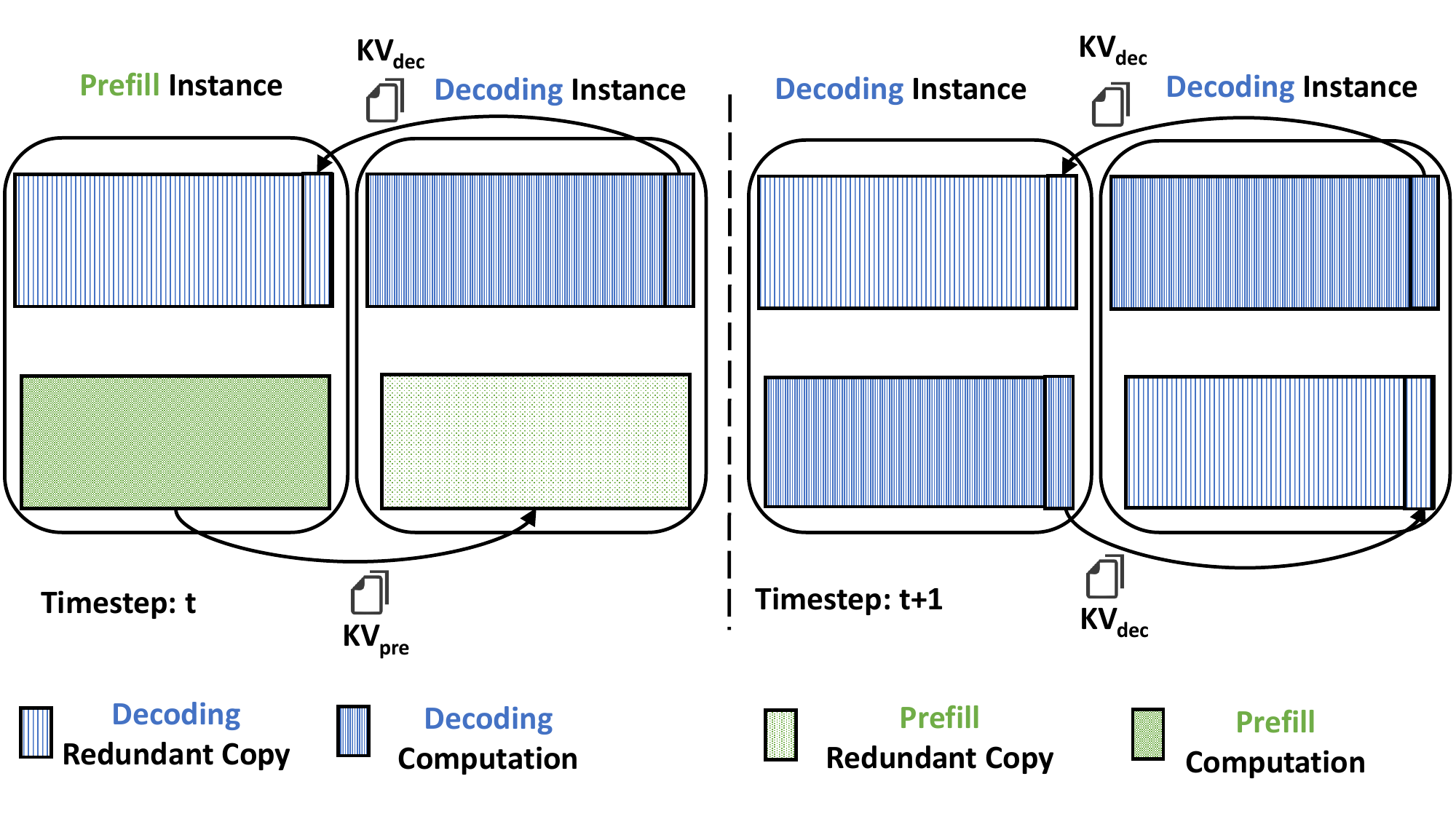}
\vspace{-0.9 cm}
\caption{
At timestep $t$, a new prefill is computed, and its newly created KV cache is transferred to the second instance. Simultaneously, decoding occurs in the second instance, and the newly computed KV cache lines are transferred to the first instance to ensure their updates. At timestep $t+1$, when there is no prefill, both instances focus solely on decoding tasks, continuously synchronizing and updating the KV caches between each other.
}
\vspace{-0.5 cm}
\label{fig:accellm_inf}
\end{figure}

\subsubsection{Dynamic Instances}

In order to tackle the peak latency problem, the prefill and decode phase is disaggregated among instances \cite{2023splitwise,2024Dejavu,2024distserve,2024Tetrinfer}. AcceLLM employs a similar strategy, however, contrary to previous \sota\ methods, where a prefill instance cannot fulfill decoding requests, and vice versa, \toolname\ supports dynamic instances capable of serving both phases, avoiding the peak latency issues by not serving them at the same time.
By ensuring that no instances remain idle during computations, this approach achieves faster request latency (TTFT, TBT, JCT). Depending on the workload volume and the number of requests processed during the decode phase, instances can dynamically switch between decoding and prefilling roles, thereby optimizing system performance.

The scheduling manager within AcceLLM disaggregates the prefill and decode phase and dynamically assesses, depending on the incoming workload rate, whether an instance will compute prefill or decoding requests. This capability enables the system to better accommodate varying rates of incoming requests. Nevertheless, transitioning an instance from one phase of computation to the other is challenging, due to the large amount of complete KV cache transfers.

\subsubsection{Redundant KV Caches}

To overcome the hurdle of frequently moving complete KV caches between instances and to streamline the conversion of instances into prefill or decoding roles, we introduce the redundant KV caches method within AcceLLM, as seen in \cref{fig:accellm_inf}.

During the prefill phase in AcceLLM, when a request is served from an instance, the entire generated KV cache is transferred to the instance designated for token generation \cite{2024distserve,2024Tetrinfer}. Unlike previous \sota\ methods, AcceLLM retains a copy of the KV cache in the instance where prefilling occurred, provided there is adequate free space in the instance's memory. Additionally, the instance responsible for the decoding phase, during its processing sends newly computed KV cache lines back to the instance storing the KV cache copy where initially the prefill was computed, ensuring its continuous updating.

Once there are no prefill requests to be served in the instance originally designated for prefilling, it can be repurposed for decoding tasks. The updated KV caches stored there are then utilized for token generation, distributing the decoding workload across multiple instances to expedite token generation and avoiding this way keeping a few of them idle. This way, there is no need for full KV caches copies before transitioning a prefill to a decoding instance, which can result into high peak latency.

However, the newly created KV caches in the former prefill instance must be transferred back to the instance where decoding initially commenced. This is necessary to ensure that an instance remains available for prefilling when new requests arrive, without stalling requests or waiting for all the requests in the decode phase to complete. Furthermore, it ensures that requests in the decode phase are not delayed, as their processing can continue in the instance where decoding initially began.

When the system consists of two instances, redundancy of KV caches for nearly all served requests can be attained. This is because, for almost every request, after its prefill is served in one instance and its decoding in another, an updated KV cache copy is retained in both instances. The only scenario where a redundant copy cannot be maintained is when the KV cache has expanded to the extent that storage in a single decoding instance is insufficient, necessitating the involvement of additional instances for service. 

In systems with more than two instances, half of the instances are allocated to either prefill or decoding tasks based on workload demands. As request rates rise and KV cache memory demands grow, some instances may replace redundant KV copies with new requests, causing a sequential shift into decoding roles. For example, in a four-instance setup lacking sufficient memory for redundancy, one instance may alternate between prefill and decode tasks, while the other three focus solely on decoding.

\begin{figure}[t]
\centering
\includegraphics[width=\linewidth]{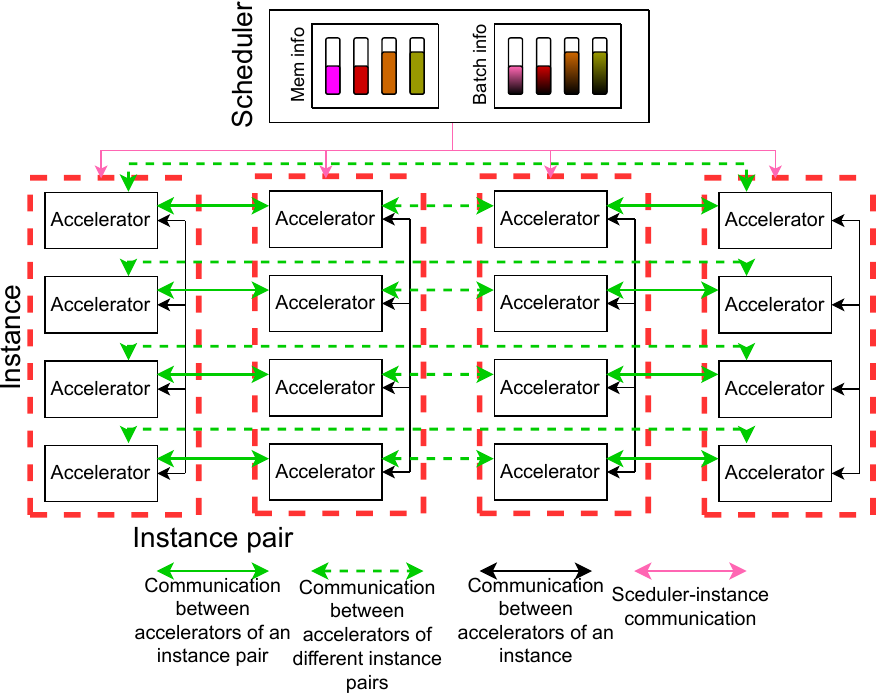}
\vspace{-0.4 cm}
\caption{AcceLLM system: The instances are organized in pairs. All the instances are capable of serving prefill or decoding requests, but not both at the same time. The scheduler assigns the computation of the new prefill in the appropriate instance  based on the free memory space. Batch size info is used to balance workload in the decoding phase. 
}
\vspace{-0.5cm}
\label{fig:acceLLM_system}
\end{figure}

\subsubsection{Load Balancing} 
Batching requests during token generation in the decoding phase is essential for maximizing accelerator resource usage, reducing the memory bandwidth pressure during these otherwise bandwidth-limited tasks \cite{2024batchingSMLrecasens}. This results in the preservation of a low TBT \cite{2024distserve}. However, excessive batching of requests leads to an increase in the TBT. As also emphasized in \cite{2024Tetrinfer,2024batchingSMLrecasens}, it is equally important to balance the number of requests per batch per instance as much as possible, especially when multiple instances are available. With the EOS token appearing at varying iteration numbers for each request, load balancing between instances becomes even more unpredictable.

Additionally, it is crucial to ensure that the total length of requests within each batch is approximately equal. This is because, during the decoding phase, the only computation that varies between requests is the attention mechanism \cite{2022daoflashattention,2023daoflashattention2,2024hongflashdecoding}, which depends on the length of the KV cache. 

With the assistance of redundant KV cache copies, AcceLLM achieves an effective balancing of the number of requests per batch among instances and evenly distributes requests according to their lengths. When a pair of instances function as decoding instances, AcceLLM ensures that the number of requests batched in each instance is approximately equal (assuming sufficient memory space). This approach leads to a reduction in TBT without necessitating additional movements of complete KV caches, while maximizing the utilization of available hardware resources. 
To balance the workload between pairs of instances, additional KV cache copies are generated gradually over time, provided that the communication bandwidth between instances isn't already saturated and there's no increase in latency overhead during token generation.

\subsection{
Technical Implementation
}

The high-level overview of AcceLLM is depicted in \cref{fig:acceLLM_system}. The system consists of instances grouped in pairs, capable of handling either prefill or decoding requests depending on the workload, but not both simultaneously. Every instance includes a full LLM model replication.
 
 \subsubsection{Instance Pair} \label{sub:pair}

As shown in \cref{fig:acceLLM_system}, the instances in AcceLLM are grouped in pairs for these experiments. 
With sufficient memory in each instance, nearly all request KV caches are replicated across the two paired instances. This setup enables one instance to handle prefills for new requests while the other continues decoding tokens for existing requests. During periods with no new requests, both instances switch to decoding, ensuring no idle time and allowing smooth transitions between operations with minimal latency overhead. When both instances are decoding, near-perfect load balancing is achieved, as redundant KV caches allow flexible batch size adjustments without latency penalties, even if imbalances arise due to completed requests.

\begin{table}[h] 
\caption{Accelerator Device Specifications} \label{tab:accelinfo}
\centering
\resizebox{\linewidth}{!}{%
\begin{tabular}{l r r r r}
     \toprule
     Device & fp16-TFLOPS & HBM cap. & HBM BW & local conn. \\
     \midrule
     910B2      & 400 & 64\,GB &  1.8\,TB/s & 392\,GB/s \\ 
     H100       & 989 & 80\,GB & 3.35\,TB/s & 900\,GB/s \\
     \bottomrule
\end{tabular}%
}
\end{table}

\begin{table}[h] 
\caption{Workload Characteristics} \label{tab:workload}
\centering
\begin{tabular}{l rrr} 
 \toprule
 Workload & Prefill & Decoding & Mean Value \\  
 \midrule
 Light & 20-500     & 20-500    &  250  \\
 Mixed & 20-1000    & 20-1000   &  500  \\ 
 Heavy & 500-1000   & 500-1000  &  750  \\  
 \bottomrule
\end{tabular}
\end{table}

\subsubsection{Scheduling Manager}

The scheduling manager in AcceLLM directs requests and handles instance conversions. When a new request arrives, it routes it to one instance in a pair, switching it to prefill computation while flagging the second instance to continue token generation for all stored requests including the redundant ones. As new KV caches are created, they are flagged for replication in the second instance, and redundant KV lines are updated. After prefill, the first instance returns to decoding, with the scheduler balancing workload by equalizing batch size and request length. Among available pairs, the one with the most free space handles the next prefill. If multiple requests arrive simultaneously, they are evenly distributed among pairs to minimize TTFT. To address inter-pair imbalances, the manager incrementally builds redundancy across pairs, given sufficient interconnect bandwidth and memory capacity.

\subsubsection{Instance}

Each AcceLLM instance comprises four accelerators and the open-source vLLM library (version 0.4.2) \cite{2023Pageattention} is utilized to manage inference requests within the instance. Tensor parallelism is configured to four, and batching is employed to minimize redundant memory reads of model weights and to enhance the utilization of available accelerator computational resources \cite{2022Orca,2022daoflashattention}. Within each instance, the vLLM code has been modified to prevent the batching of prefill and decoding operations at any point ensuring that no latency spikes occur during the decoding phase.

\begin{figure}
  \centering
  \includegraphics[width=0.7\linewidth]{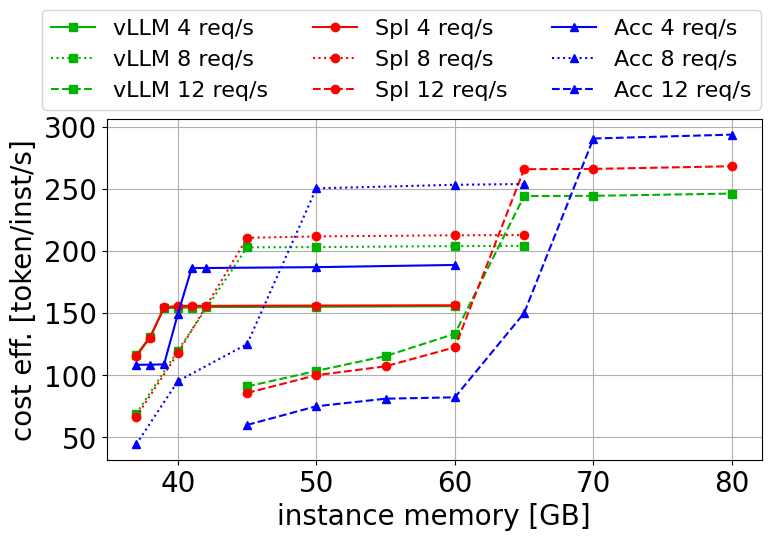}
  \label{fig:memvscost}
{\includegraphics[width=0.7\linewidth]{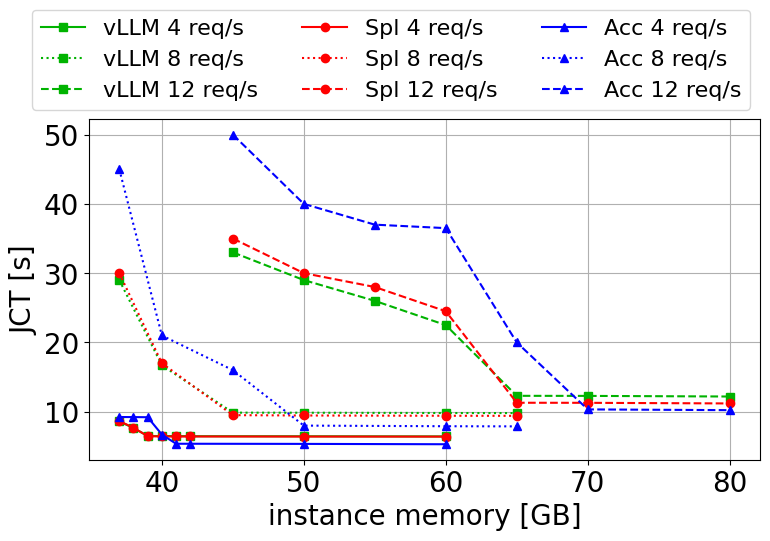}\label{fig:intevscost}}
  \vspace{-0.4cm}
  \caption{Memory requirements for mixed workload for a cluster consisting of four instances.}
  \vspace{-0.5cm}
  \label{fig:memac}
\end{figure}

\begin{figure}
  \centering
  \includegraphics[width=0.48\linewidth]{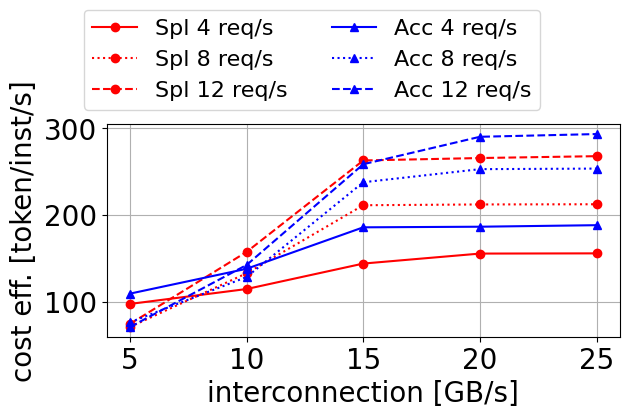}
  \label{fig:bandvscost}
  \hfill
 \includegraphics[width=0.48\linewidth]{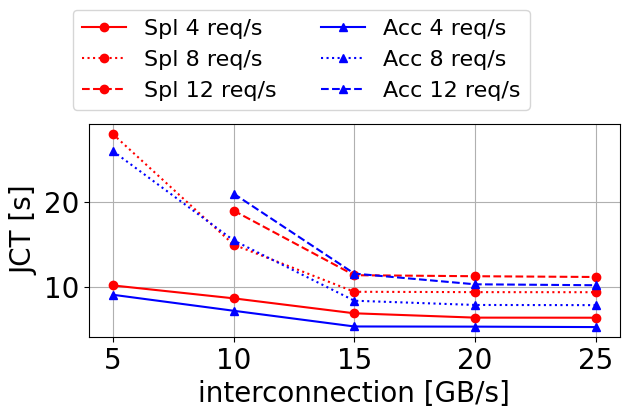}
  \label{fig:bandvsjct}
  \vspace{-0.5cm}
  \caption{Interconnection requirements for mixed workload for a cluster consisting of 4 instances.}
  \vspace{-0.7cm}
  \label{fig:bandac}
\end{figure}

\begin{figure*}
  \centering
  \begin{subfigure}
  {\includegraphics[width=1.62in, height=1.22in]
{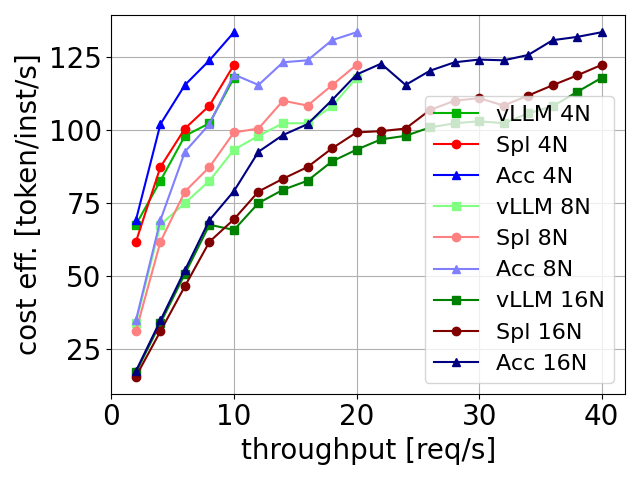}\label{Costhanamixed}}
  \end{subfigure}
   \begin{subfigure}
  {\includegraphics[width=1.62in, height=1.22in]
{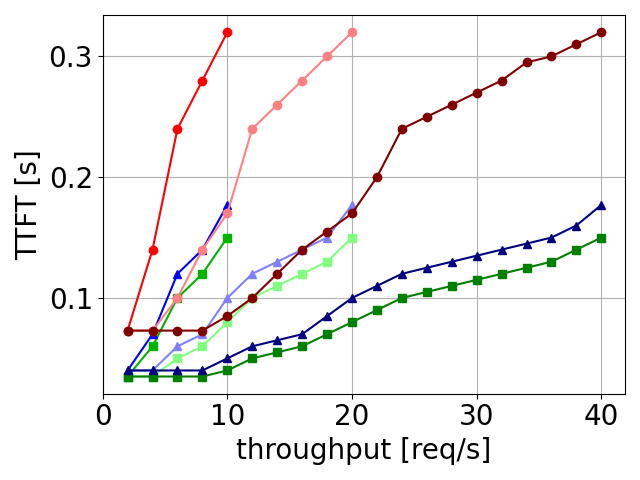}\label{ttftmmixed}}
\end{subfigure}
   \begin{subfigure}
  {\includegraphics[width=1.62in, height=1.22in]{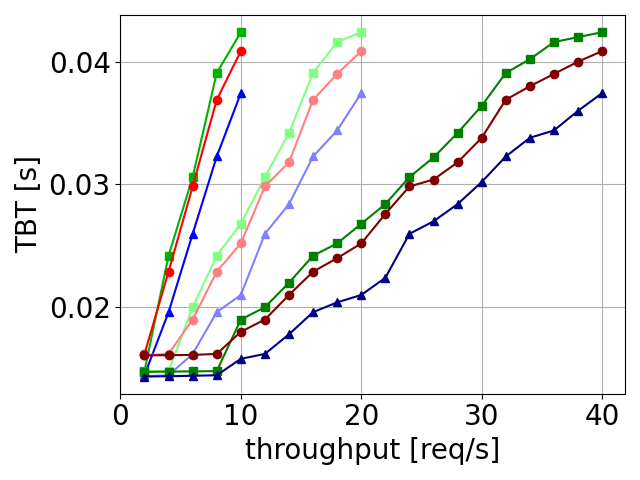}\label{tbtmmixed}}
  \end{subfigure}
   \begin{subfigure}
  {\includegraphics[width=1.62in, height=1.22in]{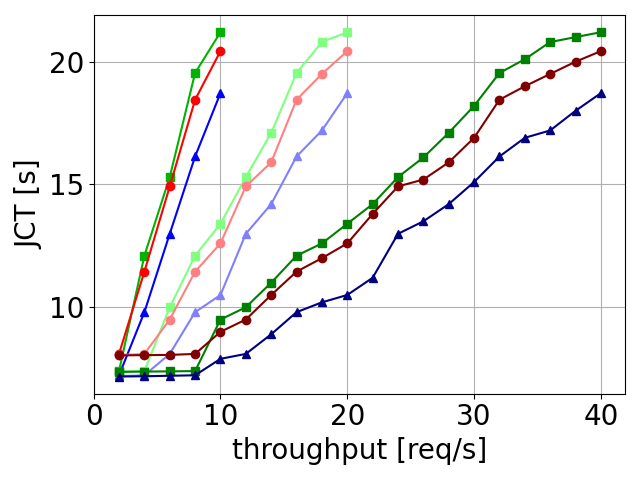}\label{jctmmixed}}
  \end{subfigure}
  \vspace{-0.4cm}
  \caption{Latency and cost efficiency for the mixed workload with simulated Nvidia H100 instances.}
  \vspace{-0.3cm}
  \label{fig:latlsnmixed} 
\end{figure*}

\begin{figure*}
  \centering
   \begin{subfigure}
  {\includegraphics[width=1.63in, height=1.22in]{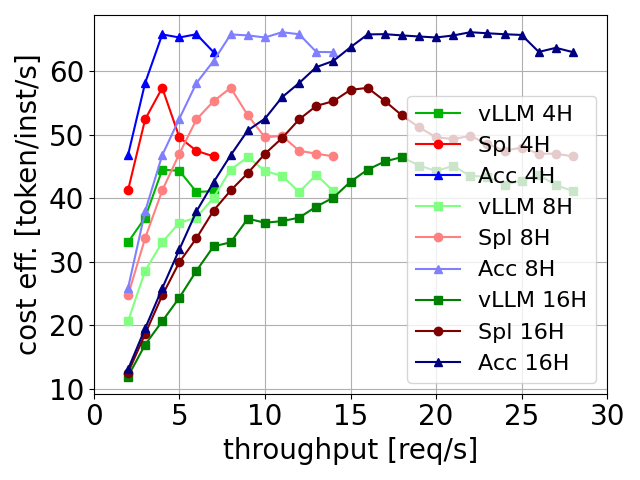}\label{costhanamixed}}
  \end{subfigure}
   \begin{subfigure}
  {\includegraphics[width=1.63in, height=1.22in]
{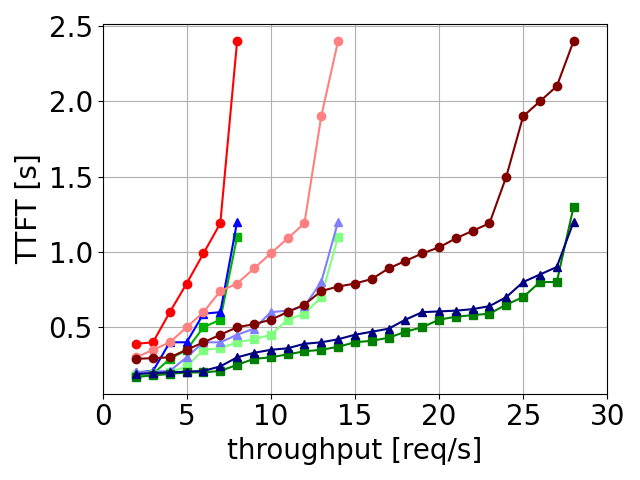}\label{ttfthmmixed}}
\end{subfigure}
   \begin{subfigure}
  {\includegraphics[width=1.63in, height=1.22in]{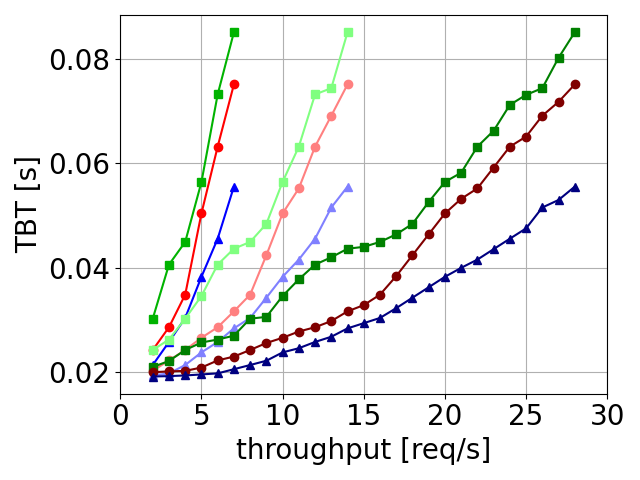}\label{TBThanamixed}}
  \end{subfigure}
   \begin{subfigure}
  {\includegraphics[width=1.63in, height=1.22in]{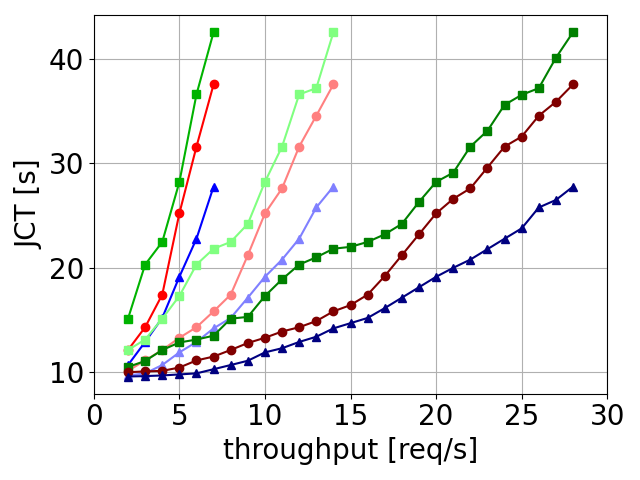}\label{JCThanamixed}}
  \end{subfigure}
  \vspace{-0.5cm}
  \caption{Latency results for the mixed workload with Ascend 910B2 instances.}
  \vspace{-0.2cm}
  \label{fig:latlshmixed} 
\end{figure*}

\subsubsection{KV Cache Transfer}

The NVlink connection between instance accelerators enables efficient KV cache migration, essential for phased disaggregated computation. Instead of waiting until the prefill phase ends which would lead to very high serving latency, KV caches are transferred per layer as they are generated, reducing JCT latency, especially for large prompts. During each layer in the prompt and decoding phases, KV caches are transferred to the paired instance, allowing computation to proceed seamlessly. The vLLM code is modified to enable these per-layer transfers and to concatenate request tensors, ensuring minimal latency overhead due to NVlink’s high-speed interconnect.

\subsubsection{Insufficient Memory Capacity for Redundancy}

When high request volumes prevent full data replication across instances, one instance is designated to handle both prefill and decoding (though not simultaneously), while the other three operate in decoding-only mode. These three instances overwrite redundant KV caches with new decoding requests, thus serving a greater number of requests. The dual-phase instance retains about one-third of the KV caches of each decoding instance as redundant copies. When no new requests are incoming, decoding is distributed across all four instances, improving load balancing.

\section{Evaluation}\label{sec:exp}

\subsection{Inference System Modeling}

Assessing AcceLLM in an actual cluster environment requires significant resource allocation. Therefore, to evaluate AcceLLM's performance on a cluster level design, we have developed a simulator that faithfully simulates the computation, HBM bandwidth, memory requirements and KV cache transfer costs of our LLM inference system. For the performance model, we leverage extensive data from experiments performed on the Huawei Ascend 910B2 accelerator and simulated Nvidia's H100 SXM5 GPU whose device specifications are shown in \cref{tab:accelinfo}.

\subsection{Workload-Baseline}

We perform our evaluations based on the Llama-2 70B parameter model \cite{2023llama2touvron}. For evaluating the experiments, three distinct workloads were tested. The token number for these workloads is shown in \cref{tab:workload}, with every request drawn from a uniform distribution.

The first baseline for comparing \toolname~'s performance is Splitwise \cite{2023splitwise}, which also applies a \sota\ disaggregation policy between prefill and decoding phases, similar to \cite{2024distserve,2024Dejavu,2024Tetrinfer}. Splitwise typically allocates most instances to prefill, though most compute time is spent on token generation \cite{2024melange,2024aladdin}. In our simulations, we prioritize decoding for Splitwise, batching both phases under high loads, and exclude non-disaggregated instances to avoid latency spikes. 

Our second \sota\ baseline is vLLM \cite{2023Pageattention}, which uses sequence parallelism for memory efficiency. We simulate Splitwise and vLLM with 4, 8, and 16 instances, applying the same inter-accelerator optimizations as AcceLLM. For Splitwise, 1, 2, and 4 instances handle prefill in the 4, 8, and 16-instance cases, respectively, while the rest focus on the more 
compute-intensive token generation phase.

\begin{figure*}[h]
  \centering
   \begin{subfigure}
  {\includegraphics[width=1.63in, height=1.22in]{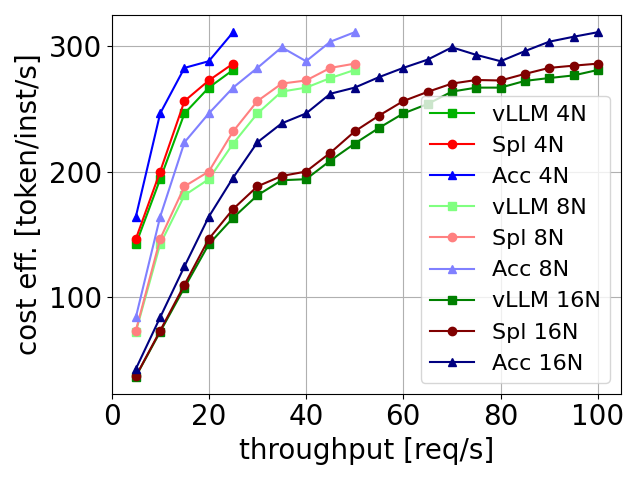}\label{costlight_nv}}
  \end{subfigure}
   \begin{subfigure}
  {\includegraphics[width=1.63in, height=1.22in]{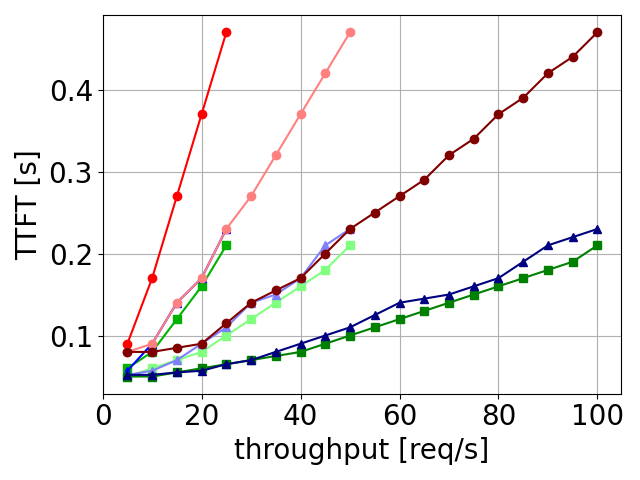}\label{ttftlight_nv}}
  \end{subfigure}
   \begin{subfigure}
  {\includegraphics[width=1.63in, height=1.22in]{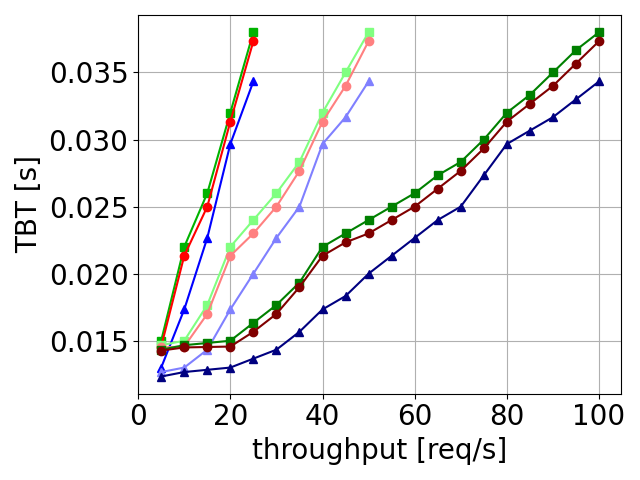}\label{tbtlight_nv}}
  \end{subfigure}
   \begin{subfigure}
  {\includegraphics[width=1.63in, height=1.22in]{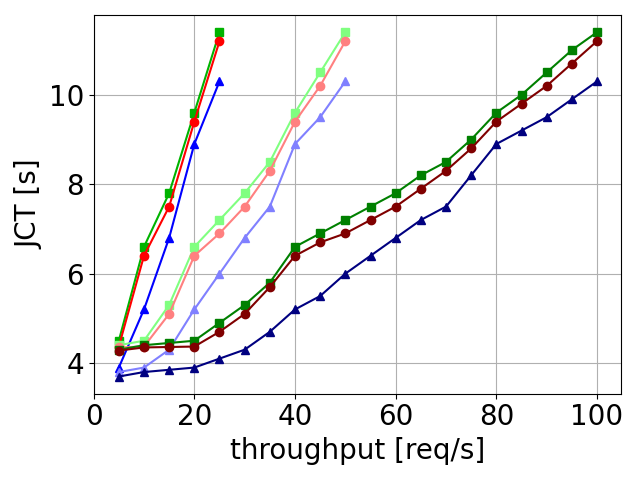}\label{jctlight_nv}}
  \end{subfigure}
  \vspace{-0.7cm}
  \caption{Latency results for the light workload with simulated Nvidia H100 instances.} 
  \vspace{-0.4cm}
  \label{fig:latlshn} 
\end{figure*}

\begin{figure*}[h]
  \centering
   \begin{subfigure}
  {\includegraphics[width=1.63in, height=1.22in]{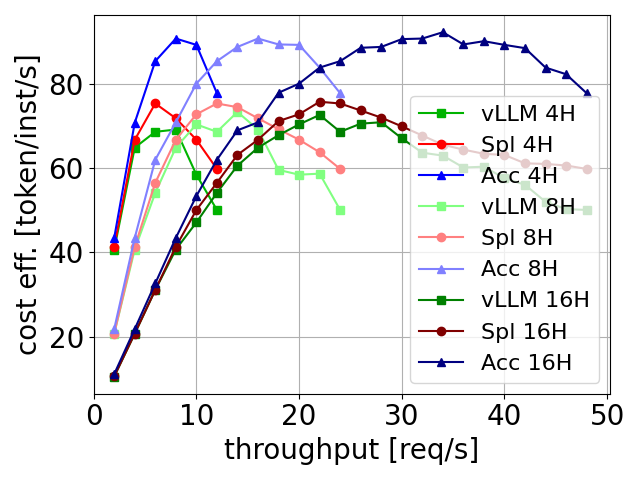}\label{costhlight_hw}}
  \end{subfigure}
   \begin{subfigure}
  {\includegraphics[width=1.63in, height=1.22in]{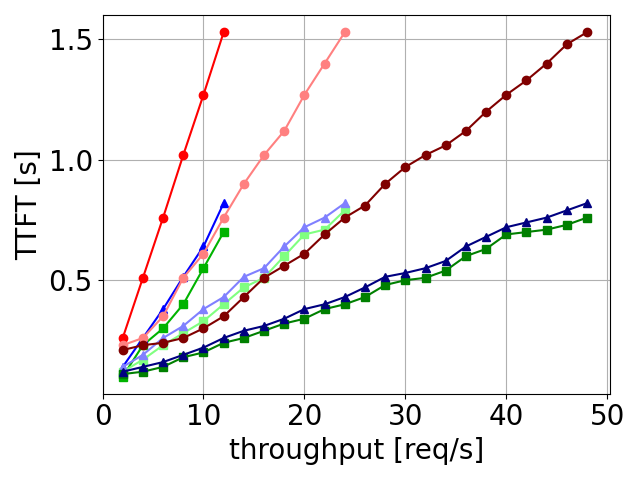}\label{ttftllight_hw}}
  \end{subfigure}
   \begin{subfigure}
  {\includegraphics[width=1.63in, height=1.22in]{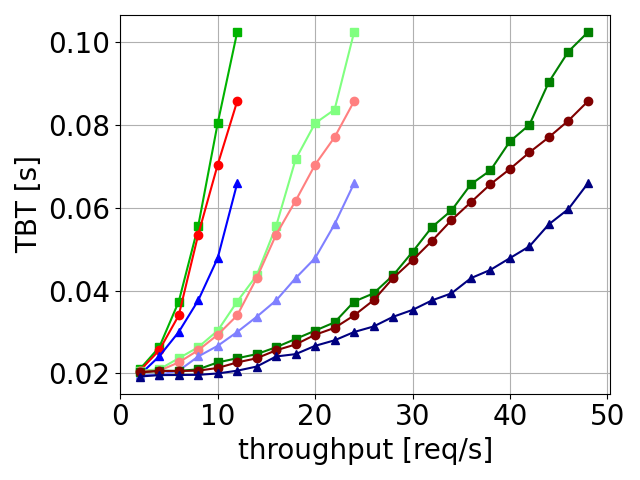}\label{tbtlight_hw}}
  \end{subfigure}
   \begin{subfigure}
  {\includegraphics[width=1.63in, height=1.22in]{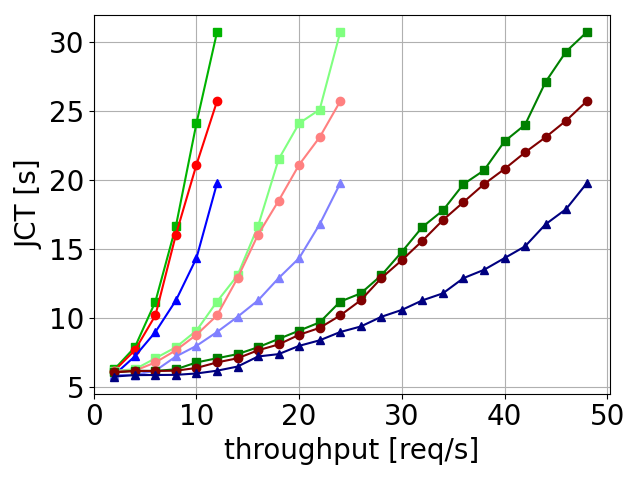}\label{jcthlight_hw}}
  \end{subfigure}
  \vspace{-0.7cm}
  \caption{Latency results for the light workload with Ascend 910B2.}
  \vspace{-0.4cm}
  \label{fig:latlshh} 
\end{figure*}

\subsection{Experimental Results}

\subsubsection*{Impact of Device Memory Capacity} Creating KV cache replicas for serving requests requires additional memory space in instances compared to other \sota\ ~methods. \Cref{fig:memac} illustrates the amount of memory needed per instance for \toolname{}~, Splitwise and vLLM in order to reach peak performance in terms of cost efficiency and JCT. It demonstrates that \toolname{} requires 1, 3.5, and 5\,GB more memory than Splitwise and vLLM for handling 4, 8, and 12 requests per second, respectively, to achieve superior performance. However, in these three cases, instance memory is never saturated in either system leaving enough space for redundancy. Additionally, in many serving systems, certain JCT requirements have to be met, meaning that the instances' memory is never fully utilized, leaving sufficient space for the redundant KV cache copies \cite{2024andes,2024melange}.

\subsubsection*{Impact of Interconnect Bandwidth} Moving KV caches between instances requires a certain interconnect bandwidth to achieve peak performance. \Cref{fig:bandac} illustrates the impact of interconnect speed on token throughput and JCT time. Both \toolname{} and Splitwise reach peak performance at approximately the same interconnect speed. The primary bulk of KV cache movement occurs during the prefill computation. In \toolname, additional KV cache line movements occur during decoding to keep the replicated caches updated. These extra movements are minimal compared to those during prefill and have a limited effect on overall interconnect utilization. We exclude vLLM from the figure because its system does not require KV cache transfers between instances.

\begin{figure*}[h]
  \centering
   \begin{subfigure}
  {\includegraphics[width=1.63in, height=1.22in]{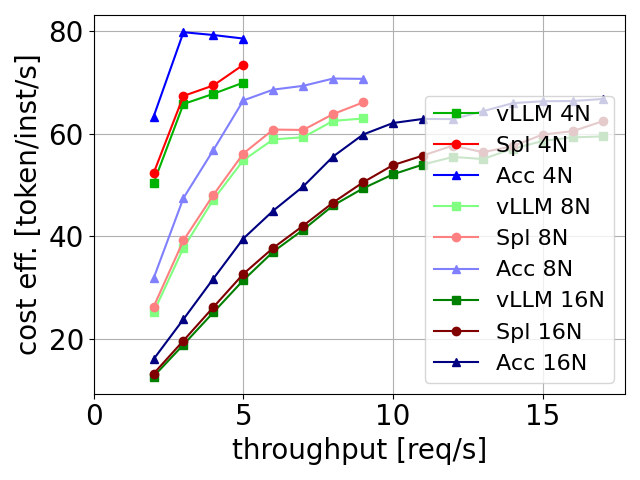}\label{heavycostnvidia}}
  \end{subfigure}
   \begin{subfigure}
  {\includegraphics[width=1.63in, height=1.22in]{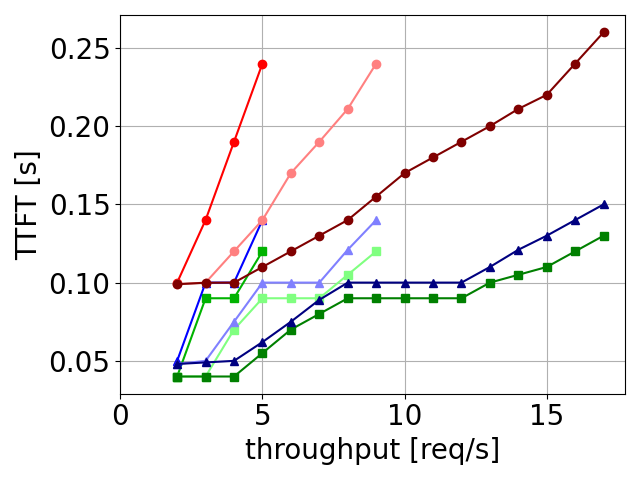}\label{heavyttftnvidia}}
  \end{subfigure}
   \begin{subfigure}
  {\includegraphics[width=1.63in, height=1.22in]{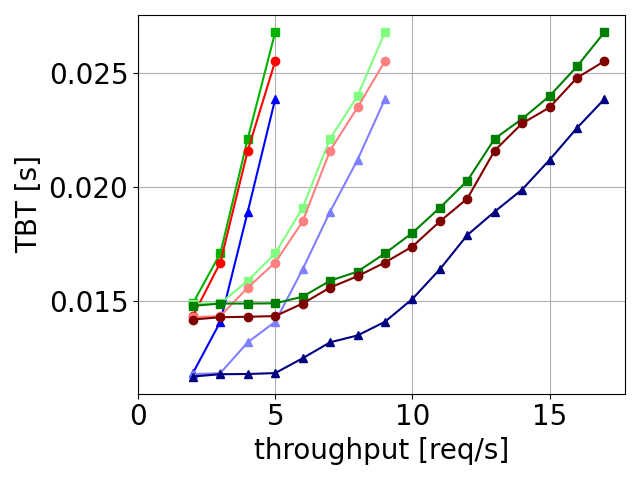}\label{heavytbtnvidia}}
  \end{subfigure}
   \begin{subfigure}
  {\includegraphics[width=1.63in, height=1.22in]{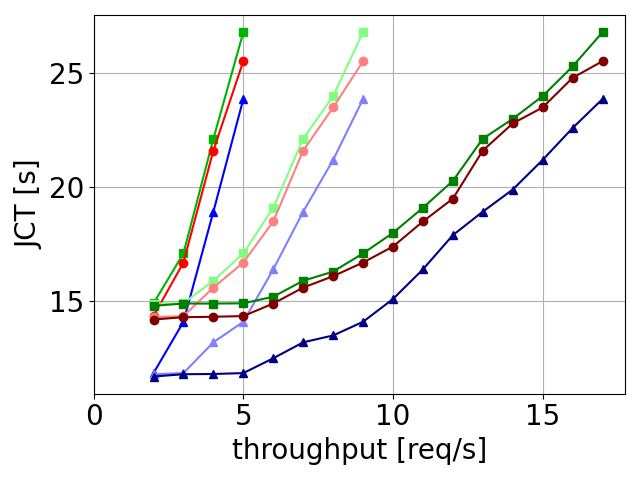}\label{heavyjctnvidia}}
  \end{subfigure}
  \vspace{-0.7cm}
  \caption{Latency results for the heavy workload with H100 instances.}
  \label{fig:heavynvidia} 
  \vspace{-0.2cm}
\end{figure*}

\subsubsection*{AcceLLM Cost Efficiency}

Considering the mixed workload case, in \cref{Costhanamixed} and \cref{costhanamixed}, we observe that across all clusters, the amount of tokens generated per instance per second follows a linear increase until it saturates and then declines. Initially, the linear growth is attributed to the positive impact of increasing the batch size on token generation. However, beyond a certain point, clusters struggle to manage this increased batch size resulting from the rise in incoming requests per second. Consequently, there is a subsequent decline in performance as the clusters' computation capacity is already saturated. 

\subsubsection*{Load Balancing With Varying Request Volumes} 
AcceLLM demonstrates superior capability in handling a larger volume of incoming requests per second before reaching computational resource saturation (\cref{Costhanamixed} and \cref{costhanamixed}), compared to Splitwise and vLLM. It outperforms Splitwise and vLLM by nearly 30$\%$ in terms of the amount of tokens generated per instance per second, resulting in up to a 30\% reduction in JCT (\cref{jctmmixed} and \cref{JCThanamixed}). This improvement is attributed to AcceLLM's dynamic load balancing of requests across instances during the decoding phase, facilitated by redundant KV caches. With the redundant KV caches, AcceLLM evenly distributes batched requests per instance regardless of their token generation length. Additionally, this reduction in JCT comes from the seamless conversion of instances from the prefill phase to the decoding one and vice versa, ensuring no instance remains idle at any given time. Conversely, in Splitwise, dedicating a large number of instances to prefill tasks often results in significant periods of resource idleness when there are no incoming requests, leading to resource wastage and a significant latency increase in JCT compared to AcceLLM.

Load balancing and the use of dynamic instances also contribute to a reduced TBT (\cref{tbtmmixed} and \cref{TBThanamixed}). These approaches allows more instances to share the computational load, ensuring a nearly equal distribution of workload across resources. In the prefill phase, AcceLLM achieves prompt computation in almost half the time compared to Splitwise (\cref{ttftmmixed} and \cref{ttfthmmixed}), facilitated by its ability to dynamically allocate double the number of instances for prefill tasks when necessary. Moreover, it computes the prefill phase in about the same time as in vLLM which batches together prefill and decoding computation.

It is worth noting that in \cref{ttfthmmixed}, when the incoming request rate increases substantially (6, 12, and 24 requests per second for systems with 4, 8, and 16 instances respectively), Splitwise queues incoming requests before prefill computation because the previous ones have not completed their prefill computation, resulting in additional latency in serving requests. This issue is not observed in AcceLLM, as more instances can be allocated dynamically for prefill when needed. 
Allocating more instances to prefill in Splitwise could reduce queuing effects but would negatively impact JCT, as fewer instances would be available for the more time-intensive decoding phase.

\subsubsection*{Load Imbalance in Previous Work} 
Similar to the mixed workload scenario, AcceLLM consistently outperforms Splitwise and vLLM across all metrics in the light workload case, with the exception of TTFT for vLLM. Looking at the TTFT metric (\cref{ttftlight_nv}) for the system with H100 accelerators, we observe that in Splitwise, prefill computation takes nearly 0.45 seconds in the worst-case scenario. This implies that prefill instances are active for only about 0.45 seconds per second, leaving them idle for the remainder of the time. However, in our AcceLLM system, all instances are utilized continuously, without idle periods which leads to the faster serving of requests (\cref{jctlight_nv}).

\subsubsection*{Overloading Prefill Instances} 
On the other hand, in Huawei's cluster (\cref{ttftllight_hw}), significant queuing of new incoming requests occurs in the Splitwise system due to the insufficient number of instances handling prefill computation. In contrast, AcceLLM's capability to dynamically convert instances between phases helps mitigate these delays in computing new requests.

\subsubsection*{Importance of Load Balancing for Heavy Workloads}  
\cref{fig:heavynvidia} shows results under heavy workloads, where high KV cache demand reduces the rate of new requests handled per second. Effective load balancing of decoding requests becomes crucial in these scenarios, as increased token generation results in larger batch sizes, which can lead to imbalance across instances. By employing load balancing, AcceLLM overcomes this issue, serving requests up to 3 and 3.5 seconds faster than Splitwise and vLLM, respectively (\cref{heavyjctnvidia}).

\begin{figure}
  \centering
   \begin{subfigure}
  {\includegraphics[width=1.55in, height=1.21in]{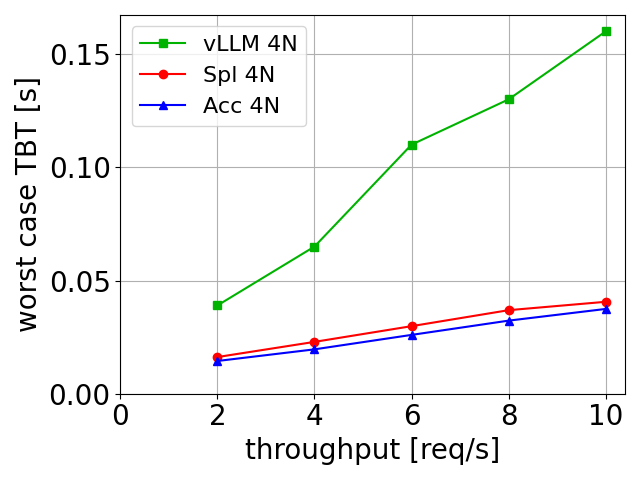}\label{fig:worstnvidia}}
  \end{subfigure}
   \begin{subfigure}
  {\includegraphics[width=1.55in, height=1.21in]{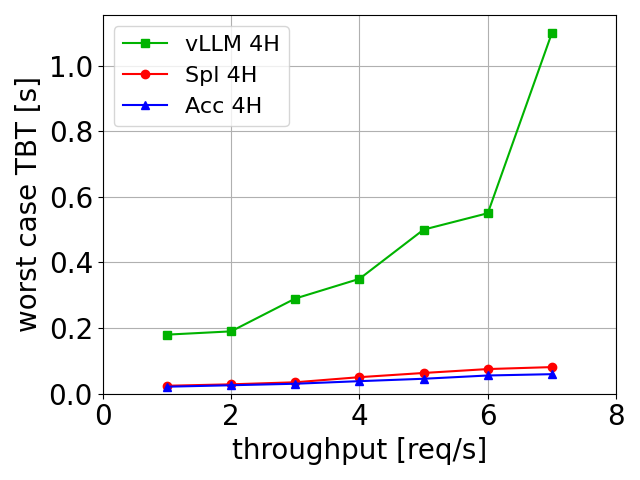}\label{fig:worsthuawei}}
  \end{subfigure}
  \vspace{-0.5cm}
  \caption{Worst case TBT latencies.}
  \vspace{-0.7cm}
  \label{fig:worsttbt}
\end{figure}

\Cref{fig:worsttbt} illustrates the worst-case scenario for TBT latency. In vLLM, the prefill and decoding processes are batched together, resulting in some tokens being generated very slowly since they are computed alongside the incoming prompt. AcceLLM prevents these latency spikes by disaggregating the prefill and decoding phases, ensuring they are computed by separate instances.

\section{Conclusions}
\label{sec:con}

We have presented \toolname, a novel methodology that employs redundant copies of LLM inference data to minimize latency and to ensure efficient allocation of hardware resources.
Our methodology leverages dynamic distinct instances and achieves load balancing. 
Targeting simulated Nvidia H100 and Huawei Ascend 910B2 accelerators, we demonstrate significant gains of up to 30\% in key user-experience metrics, sustaining negligible memory impact, compared to \sota\ methods.

\bibliographystyle{mlsys2025}
\bibliography{references}

\end{document}